\documentclass{iopart}

\usepackage{graphicx}

\begin{document}

\title[Large scale numerical modelling of laser wakefield accelerators]{Exploiting multi-scale parallelism for large scale numerical modelling of laser wakefield accelerators}

\author{R A Fonseca$^{1,2}$, J Vieira$^2$, F Fiuza $^3$, A Davidson$^4$, F S Tsung$^4$, W B Mori$^4$ and 
L O Silva$^2$}

\address{$^1$Departamento Ci\^encias e Tecnologias da Informa\c{c}\~ao, ISCTE - Instituto Universit\'{a}rio de Lisboa, 1649-026 Lisboa, Portugal}
\address{$^2$GoLP/Instituto de Plasmas e Fus\~ao Nuclear, Instituto Superior T\'ecnico, 1049-001 Lisboa, Portugal}
\address{$^3$Lawrence Livermore National Laboratory, Livermore CA, U.S.A.}
\address{$^4$University of California Los Angeles, Los Angeles CA, U.S.A.}
\ead{ricardo.fonseca@iscte.pt}

\begin{abstract}
A new generation of laser wakefield accelerators, supported by the extreme accelerating fields generated in the interaction of PW-Class lasers and underdense targets, promises the production of high quality electron beams in short distances for multiple applications. Achieving this goal will rely heavily on numerical modeling for further understanding of the underlying physics and identification of optimal regimes, but large scale modeling of these scenarios is computationally heavy and requires efficient use of state-of-the-art Petascale supercomputing systems. We discuss the main difficulties involved in running these simulations and the  new developments implemented in the OSIRIS framework to address these issues, ranging from multi-dimensional dynamic load balancing and hybrid distributed / shared memory parallelism to the vectorization of the PIC algorithm. We present the results of the OASCR Joule Metric program on the issue of large scale modeling of LWFA, demonstrating speedups of over 1 order of magnitude on the same hardware. Finally, scalability to over $\sim 10^6$ cores, and sustained performance over $\sim 2$ PFlops is demonstrated, opening the way for large scale modeling of laser wakefield accelerator scenarios.
\end{abstract}

\submitto{\PPCF}
\maketitle

\section{Introduction}

Extreme laser intensities, in excess of $10^{23} \rm{W/cm}^2$, and laser pulse durations shorter than 1 picosecond, where the electron quiver motion in the laser field becomes relativistic \cite{Mourou:2006ef}, are an extraordinary tool for new physics and new applications. Their interaction with plasmas can generate tremendous accelerating fields, which can be used to accelerate particles to high energies in very short distances in what is called laser wakefield acceleration (LWFA) \cite{Tajima:1979cb}. In LWFA an intense laser driver propagating through a plasma drives a strong, non-linear plasma wave, by pushing the plasma electrons aways through radiation pressure. The excitation of the wake is non-linear in most scenarios, and the fields within the wake structure are electromagnetic in nature requiring full electromagnetic treatment. Also, the trajectories of individual electrons can cross each other and their oscillation energy in the wakefield will reach highly relativistic values. This makes purely theoretical descriptions difficult, and requires the use of full kinetic models that track the motion of individual particles and that solve for the relativistically correct equations of motion

Numerical modeling has played a key role in the development of this field, for example predicting the generation of low energy spread electron bunches \cite{Pukhov:2002gw, Tsung:2004kv} that were later demonstrated experimentally \cite{Faure:2004hb,Geddes:2004cx,Mangles:2004ea} and that can now reach energies of $\sim 2$ GeV \cite{Wang:2013el}. Modeling these scenarios requires us to accurately model the propagation and evolution of a short ($\sim 50$ fs) intense laser driver over long distances ($\sim 1 - 100$ cm), how the wake is excited and how it evolves and how the properties of the injected beams evolve as they are accelerated. However, full 3D scale of modeling of these scenarios is computationally heavy, and require the use of high end high performance computing (HPC) systems. These systems rely on a hierarchy of parallelism that allows many systems worldwide to operate at maximum performances well in excess of 1 Petaflop \cite{top500}, following a trend that can be traced back 60 years, and that relied on substantial changes in computer architecture. While these systems provide the raw computing power required, numerical codes must evolve and adapt to match these architectures and efficiently use these resources for modeling relevant problems. 

In this paper we discuss the issues involved in exploiting this multi-scale parallelism for running efficient large scale LWFA simulation on high end HPC systems. We begin by giving some details on the numerical algorithm used, and estimate the computational cost of running these large scale numerical experiments of LWFA. We follow by describing the multi-scale parallel hierarchy that represents the most common paradigm of modern HPC systems. We then discuss the techniques employed for scaling the code to $10^5 - 10^6$ cores systems, and for exploiting existing hardware acceleration features. We continue by discussing the results of the OASCR Joule Software effectiveness metric test for modeling LWFA on a full system, and show results for the code performance on Tier-0 systems. We conclude with an overview.

The developments presented in this paper were done in the OSIRIS framework \cite{Fonseca:2002ha}, which is a massively parallel, fully relativistic, electromagnetic particle-in-cell code that was originally developed for the study of plasma based accelerators which is, to the best of our knowledge, the only code currently supporting all of these features. There are several other codes that have been applied to 3D modeling of LWFA such as VLPL \cite{Pukhov:1999wg}, CALDER \cite{Lefebvre:2003kl}, Vorpal \cite{Nieter:2004ea}, and others \cite{Ruhl:2006tw, Geissler:2006bd, Bowers:2008bu}, and some implement some of the features described here, such as vectorization of the PIC algorithm \cite{Bowers:2008bu}. There has also been significant developments in implementing these algorithms in novel architectures, most notably using GPUs \cite{Burau:2010kx}.

\section{Numerical modeling of LWFA}

Due to the highly non-linear kinetic processes occurring in high intensity laser plasma interactions, and in particular in laser wakefield acceleration scenarios, kinetic modeling is required to capture the full detail of the acceleration and injection mechanisms in LWFA. The most complete method that can presently be employed for modeling these scenarios is the fully explicit, full Maxwell solver, fully relativistic, particle-in-cell (PIC) method \cite{Dawson:1983wr}. In this method, plasma particles are represented by a set of macro-particles that interact through forces deposited on a grid, implementing a particle-mesh algorithm. These particles are initialized with positions within a grid and momenta. At each timestep, the particle motion generates an electrical current that is deposited onto the appropriate points on the grid. MaxwellÕs equations are then used to advance the electric and magnetic fields, generally through the finite difference time domain (FDTD) method. The updated fields can then be interpolated at the particle positions to calculate the force acting on the particle and then advance the fully relativistic equations of motion for the particle, thus closing the simulation loop. The grid cell size is chosen to resolve the smallest scale length of physical relevance, typically the laser wavelength in LWFA scenarios, and the time step is then chosen according to the Courant condition.

If we apply this algorithm to a 10 GeV LWFA stage, theoretical scalings \cite{Lu:2007eb} indicate that this will require an acceleration on the order of  $\sim 0.5$ m, using a plasma density of $10^{17} \rm{cm}^{-3}$. The simulation needs to resolve the laser wavelength $\lambda_0$ longitudinally along the laser propagation axis with $\sim 20$ points per $\lambda_0$, and transversely the plasma skin depth $c/\omega_{pe}$ with $\sim 10$ points per $c/\omega_{pe}$. The grid cells will be much smaller along the longitudinal direction, which will improve the numerical dispersion of the FDTD method along the laser propagation axis. The algorithm uses a moving window that follows the laser as it propagates in the plasma, that should be large enough to model $\sim 20$ skin depths longitudinally (or 2 accelerating buckets), and $\sim 6$ laser spot sizes, $w_0$, transversely. For a $\lambda_0 \sim 1 \rm{\mu m}$ laser focused on a $w_0 \sim 100 \rm{\mu m}$ spot the simulation grid will therefore require $\sim 10^9$ cells to model the required interaction volume in 3D. The simulation time step will be on the order of the longitudinal resolution and the total number of iterations can be calculated by dividing the total propagation distance by this value, yielding a total of $\sim 10^7$ time steps. Using 8 particles per cell, the total number of particles that need to be modeled will be on the order of $\sim 10^{10}$. A 3D EM-PIC algorithm, using linear interpolation, requires $\sim 350$ operations per particle and $\sim 80$ operations per grid cell, which leads to a total number of required calculations on the order of $\sim 10^{19}$. Realistic, high-fidelity simulations therefore require the efficient use of Petascale computing systems, and adapting our algorithms and codes to the specific requirements involved.

The high performance attained by present computing systems is reached by exploiting several levels of parallelism, ranging from inside a single compute core to a full system consisting of $\sim 10^6$ cores. The details of each of these systems will vary, but the currently prevailing paradigm generally follows a three level hierarchy of parallelism: i)  At the highest level we have a computer network of independent computing nodes, each one having private memory spaces and processing units, ii) each of these nodes is generally comprised by a set of CPUs/cores that share the memory and other resources inside these nodes, and iii) at the lowest level the computing cores have vector units capable of single instruction multiple data (SIMD) calculation i.e. being able to execute the same instruction on a set of data. Efficient use of these systems requires taking advantage of all these levels of parallelism, only then tapping the real potential of current day Petascale computing systems. Over the past decade some systems have also include some additional accelerator hardware at the node level, such as a general purpose graphics processing unit (GPGPU), or Intel Xeon Phi (MIC) cards, such as the ones used in the current number 1 machine in the world. These can also be viewed as a computing node with memory shared by a (large) number of computing elements and some of the techniques described below can be applied to these systems, but they will not be covered in detail here.

\section{Scaling to $\sim 10^5 - 10^6$ cores}

Given the computational requirements for realistic simulations PIC codes must be designed from scratch to be used in massively parallel computer systems. As described above, at the highest level these systems can be viewed as a network of independent processing elements (PEs): each PE has its own private memory space, and information residing on other nodes can only be accessed by exchanging messages over the interconnect. PIC codes are good candidates for parallelization in these systems because the particle calculations are inherently local, requiring only information close to the particle position on the grid. The field solver chosen should also be a local field solver, requiring only the information regarding neighboring cells, which requires less parallel communication than a global field solver such as a spectral solver. With this in mind, the code can simply split the problem across multiple processing elements using a spatial domain decomposition. In this parallel partition each node will handle a fraction of the global simulation grid and the particles that occupy that region of simulation space. The simulation grids use guard cells that hold values belonging to neighboring nodes, that are updated only once per iteration. These guard cells must be enough to accommodate both the needs of the field solver and the particle interpolation scheme. After advancing particles, these are checked to see if they have crossed the node boundary, and if so, they are sent to the appropriate neighboring node. This parallelization strategy is well established \cite{Wang:1995tp} and has proven to be very effective for scenarios with small density variations, as shown by the results in section \ref{sec_performance} below. 

\begin{figure}
\begin{center}
\includegraphics{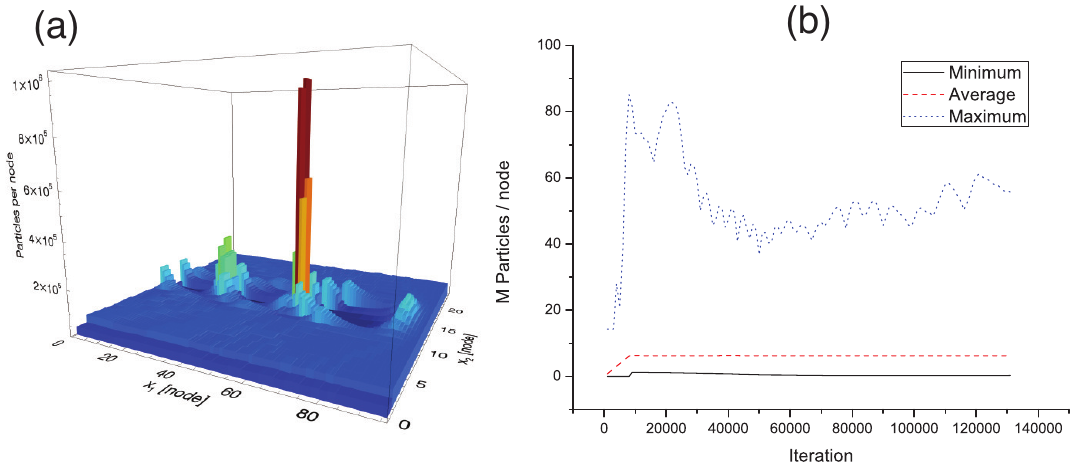}
\caption{(a) Number of particles per node at the end of 40000 iterations for a central 2D slice of the 3D parallel partition (b) Evolution of the minimum (solid), average (dashed) and maximum (dotted) number of particles per node over the whole simulation.}
\label{fig_imbalance}
\end{center}
\end{figure}

However, for high resolution 3D LWFA simulations this strategy leads to poor performance at high ($\sim 10^5$) core counts due to load imbalance. The plasma dynamics during LWFA, where the wave structure and the self-injection mechanism places many particles in a localized region of real space, leads to a severe accumulation of particles in a very narrow region at the back if each wavelength. If the region where the particles accumulate resides on a very small number of the total PEs then a severe load imbalance will arise. Figure \ref{fig_imbalance} shows results from a 3D LWFA simulation that is described in detail in section \ref{sec_oascr} below that illustrates this problem. Compared to a warm plasma benchmark run on the same partition, the code ran approximately $\sim 9$ times slower as a result of the severe load imbalance occurring in this simulation. Figure \ref{fig_imbalance}(a) plots the number of particles per core in a 2D slice of the 3D parallel partition closest to the laser propagation axis at iteration number 50000, showing a large accumulation of particles behind the laser pulse. The average load imbalance for this simulation (ratio between the maximum number of particles per node and average number of particles per node) was 8.68 which is very close to the observed slowdown.  This situation persists throughout most the simulation, as shown in figure \ref{fig_imbalance}(b). As soon as the plasma enters the simulation box it starts to accumulate behind the laser and a significant imbalance is maintained over the whole simulation length. The dynamic nature of this imbalance should also be noted, since the position of the accumulated particles in the simulation window will vary over time and slowly move between PEs. 

\subsection{Shared memory parallelism}

One way to address this issue is to explore the fact many computer cores share memory access inside a node. The parallelization strategy described previously is designed for a distributed memory system, where each core handles a separate region of simulation space. This has proven extremely efficient because at high core counts the simulation space handled by each core is quite small, and the probability that significant load imbalance will occur is quite high. However,  since many cores will share memory,  we can therefore have them share a given simulation region, whose volume will be much larger, and distribute the particles inside this larger volume evenly across these cores. The workload for these cores will always be perfectly balanced; globally some imbalance may still occur, but since the volume handled by each group of cores is much larger, the probability for significant load imbalance will be lower.

This can be achieved by parallelizing the particle pusher using a shared memory approach: each process will be responsible for a larger region of simulation space to be shared by a given number of cores. This process will spawn $n_T$ threads (ideally matching the number of cores available) to process the particles, dividing particles evenly across threads.  This may result in memory conflicts, as particles being processed by different threads may try to deposit current to the same memory location. To overcome this, we also use $n_T$ copies of the electric current grid, and each thread will only deposit to 1 of these copies. After this has finished, the algorithm will accumulate all current grids in a single current grid, also using dividing the work over $n_T$ threads. This algorithm has the overhead of having to zero multiple current grid copies and also performing the reduction operation of all them, and therefore will not scale for an arbitrary number of cores, and a balance needs to be found. However, this overhead is quite small, especially because these operations can also benefit from shared memory parallelism, and it is generally possible to scale it to all cores in a CPU with good efficiency. Since a larger simulation volume is now being shared by multiple cores, it is also necessary to use shared memory parallelism on other sections of the code that would now represent a more significant part of the simulation loop, such as the field solver. However these are trivial to parallelize in shared memory nodes using textbook strategies with OpenMP constructs. This technique has the added benefit of reducing the communication volume (the total number of cells that need to be communicated between nodes), slightly offsetting the overhead of the shared memory algorithm.

We measured the performance of the shared memory algorithm using a standard warm plasma benchmark on 2 different systems, a Cray XT5 (Jaguar/ORNL), using 2, 6-core, AMD 6276 cpus per node, and an IBM BlueGene/P (Jugene/JFZ), using 1, 4-core, PowerPC 450 cpu per node. For the baseline we used the standard distributed memory algorithm (MPI only). On the IBM system the performance degradation was negligible up to 4 threads per MPI process (the maximum) value. On the Cray system, the performance degradation was below 10\% up to 6 threads per MPI process (or 1 MPI process per cpu). When using 12 threads per MPI process the performance drops 27\% below the baseline because the additional calculations become relevant and also because of memory affinity issues.

\begin{figure}
\begin{center}
\includegraphics[width=10cm]{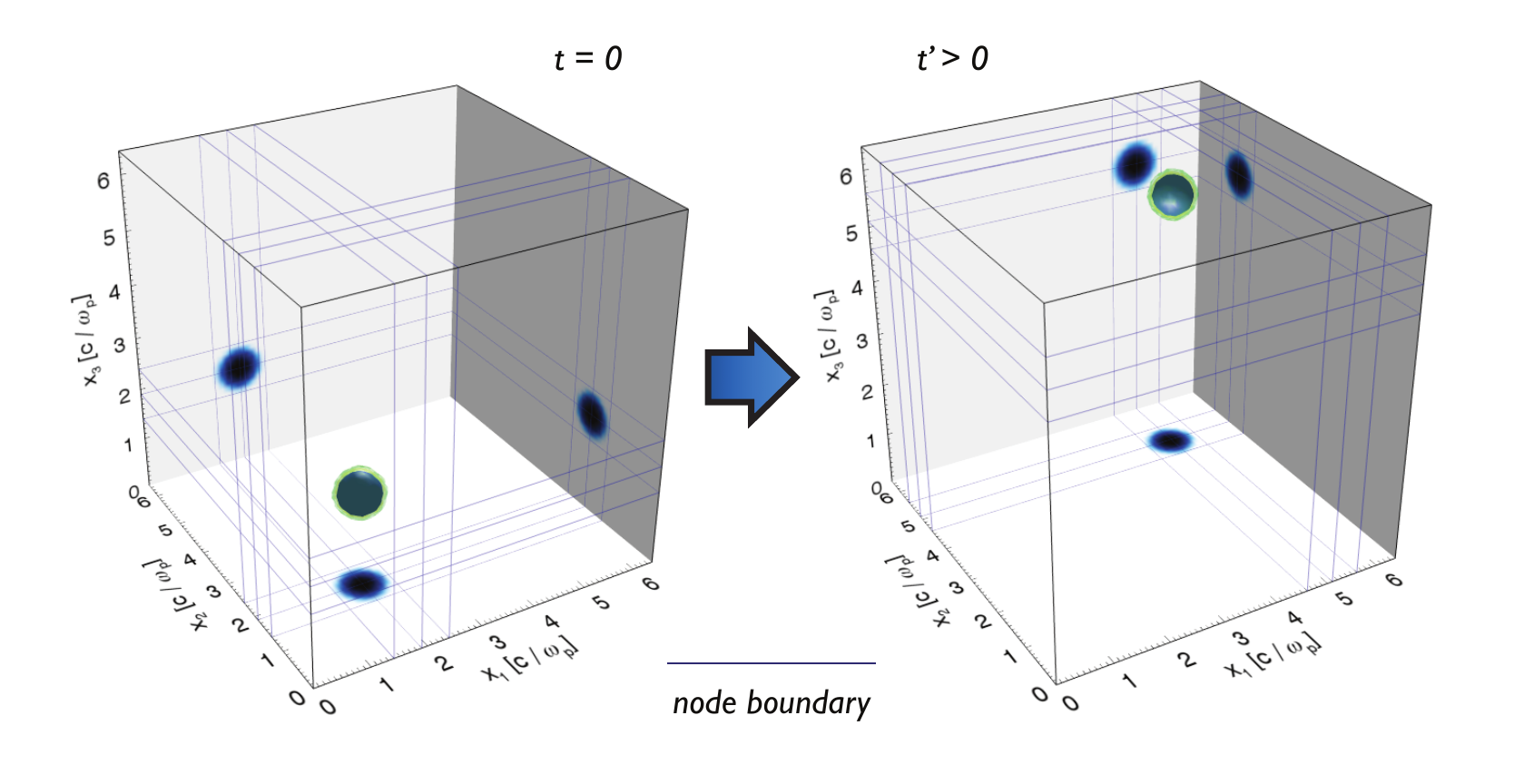}
\caption{Multidimensional load balance algorithm}
\label{fig_dyn_lb}
\end{center}
\end{figure}

\section{Multi-dimensional Dynamic Load Balance}

Another approach to the imbalance problem is to dynamically adjust the positions of node boundaries to adjust the load in each computing node, attempting to maintain an even load across processors \cite{Ferraro:1993tg, Fonseca:1134500}. The algorithm has 2 steps: i) determining the best possible partition from current computational load and ii) redistribute the domain boundaries. Both steps are non-trivial, since finding an optimal partition can be quite difficult, and redistributing the domains can represent a large overhead, with significant communication between nodes. 

The only scenario for which a single solution exists for the best parallel partition is for a one-dimensional parallel partition. At high core counts, however, one must resort to multi-dimensional partitions, and attempting to improve the load imbalance in one direction may result in worse results in the other directions. One possible solution is to assume that the computational load is a separable function (i.e. load is a product of $f_x(x) \times f_y(y) \times f_z(z)$). In this case dynamic load balance in each direction becomes independent of each other and a solution can be found as shown in figure \ref{fig_dyn_lb}. This assumption is not true for all scenarios (e.g. spherical plasma) but will generally result in a better parallel partition. It should also be noted however that at very high core counts, there isnÕt significant room to move your boundaries, given that we will be close to the minimum limit of cells per node, and therefore our chances for optimization are quite low. 

\begin{figure}
\begin{center}
\includegraphics[width=10cm]{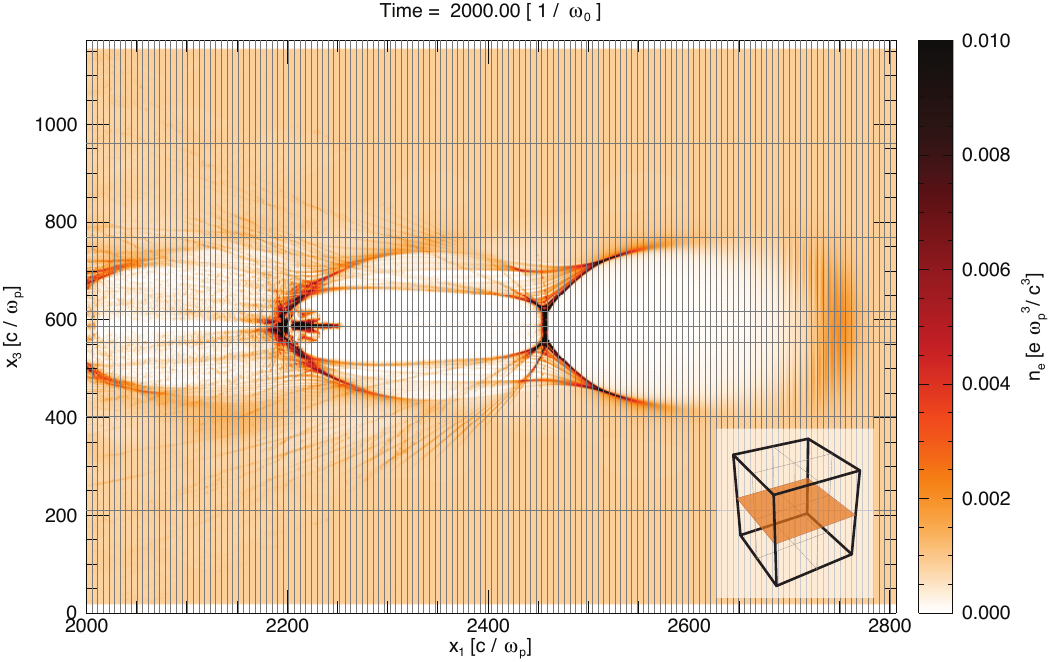}
\caption{Plasma density in a 2D slice at the center of the 3D simulation box showing the parallel node boundaries (thin grey lines)}
\label{fig_node_partition}
\end{center}
\end{figure}

Figure \ref{fig_node_partition} shows this algorithm being used in a LWFA scenario. In this example the algorithm was applied simultaneously in the $x_2$ and $x_3$ (not shown here) directions, while the partition along $x_1$ was kept constant, since there was no significant room for the boundaries to move. The algorithm concentrated more partitions near the center of the laser propagation axis where particles accumulate, and has wider partitions close to the edge of the simulation box (similar partitions were used along $x_3$). For this particular example however, although the algorithm improved the average imbalance over the simulation by 30\%, the code showed no performance improvement. This was due to the fact that the imbalance improvement was not sufficient to offset the overhead of reshaping the simulation partition.

\section{Using vector (SIMD) units}

Further improving simulation performance requires exploiting the vector capabilities available on most current CPUs. As mentioned earlier, these CPUs include a single instruction multiple data (SIMD). These vector units allow for the same instruction to be applied on a $n_V$ wide vector of distinct values at once, enabling much greater floating-point performance. The particular width of the vector will depend on the system; for example the Power A2 QPX unit (BlueGene/Q) allows for 4-wide double precision calculations and the x86 AVX unit allows for 8-wide single precision calculations. Modern optimizing compilers will automatically generate code for these units by attempting to identify vectorization opportunities in serial code, but optimal performance can only be achieved by rethinking the algorithm and implementing the vector code explicitly. Also, this is not required for all parts of the PIC loop, and optimizations should be focused on the most time consuming steps of the algorithm, in particular the particle push (field interpolation and equations of motion) and the electric current deposition that take up $\sim 90\%$ of execution time.

\subsection{Data structures for vector code}

Efficient use of the SIMD hardware requires first and foremost the ability to efficiently load the data into the vector registers for calculations. With current CPUs global memory access is generally much more expensive than computations, since up to 8 operations can be performed in a single cycle, efficiently loading the data into the vector registers is critical. Ideally, data structures should be organized so that data could be loaded into vector registers in a single instruction e.g. by grouping each position component in a different buffer sequentially in memory. However, to benefit from the existing code base that handles a large set of functionalities, ranging from initialization, to diagnostics and parallel communications, we need to maintain compatibility with the existing serial data structures, where particle data is is stored with different $x$, $y$ and $z$ position components for the same particle sequentially in memory. 

An efficient way of translating between the two data structures is to use data shuffling instructions that are also available in SIMD units. These instructions allow us to efficiently exchange vector elements inside a single vector or pairs of vectors without accessing memory, and with very little overhead. For example, when reading 3D positions for processing in a 4-wide vector unit we i) read 3 vectors sequentially from the particle buffer, that will correspond to 12 position components and 4 particles, and ii) shuffle the vector data to get one vector of 4 $x$ positions, one vector of 4 $y$ positions and one vector of 4 $z$ positions. This operation is known as a $4 \times 3$ transpose, and it can be done very efficiently using only the vector registers, giving a minimal overhead of about $\sim 3$ operations per particle (compared to the total $\sim 350$ floating point operations required for a linear push). For storing the particles back to memory, the opposite operation is performed, again allowing us to maintain the global data structures with negligible performance impact. This method can easily be extended to work with other vector widths. It should also be noted that for benefiting from the highest performance of most SIMD units available the simulation should be done in single precision, as this allows for more values to be processed in a vector of the same bit width. To ensure accuracy and stability of the simulation, this requires storing particle positions relative to the local mesh vertex, normalized to cell size, together with the particle mesh vertex index. This brings the added benefit of making the noise properties of the algorithm uniform across the entire grid, and allows large scale LWFA simulations to be performed correctly in single precision.

Regarding grid quantities, the field interpolation calculations present a different challenge, requiring what is known as a gather operation. We need the appropriate field values for the 4 particles being processed to be loaded into a single vector: since particles are free to move about the grid, these field values will belong to random positions in the grids. Furthermore, when using a Yee grid \cite{YEE:1966uw} for the fields, different field components will be defined at different positions inside the cell, and a particle may need to interpolate data from different cells for different components. While the current generation of CPUs do not support vector gathering natively, it can be implemented through a set of serial reads and shuffle operations, with negligible overhead when compared to a serial read, again maintaining compatibility with the global memory structures. For the electric current deposition a different approach was chosen: we add a dummy component to the electric current so that there are 4 values per cell. This allows us to very efficiently accumulate the current in a given cell reading these 4 values into a vector at once, vectorially adding the new current and then storing the 4 values also in a single instruction. Besides adding a small memory overhead, this  has no impact in the global code structure.

\subsection{Vector version of the PIC algorithm}

PIC codes, and the particle pusher/current deposition in particular, are good candidates for vectorization since most operations acting on each particle are exactly the same and independent from each other. The exception is the electric current accumulation on the global grid, where more than 1 particle in the same vector may be accumulating current in the same cell, which requires special treatment to avoid memory collisions. The vectorization strategy for an $n_V$-wide vector unit is therefore straightforward: i) we load $n_V$ particle data (position and momenta) into the vector unit, using the procedure described in the previous section, and calculate the interpolation weights, ii) we interpolate the EM fields for these $n_V$ particles, loading the appropriate field values and using the previously calculated weights, iii) we advance the $n_V$ particles equations of motion using the particle data and the interpolated fields, iv) we create up to $4 \times n_V$ virtual particles trajectories for current deposition and v) we store the $n_V$ particles back to main memory. With the exception of the trajectory splitting, which is mostly a serial operation, all calculations are done vectorially. 

We then turn our attention to depositing the current using a charge conserving method \cite{Villasenor:1992uk,Esirkepov:2001ve} of the virtual particles trajectories created and i) load $n_V$ virtual trajectories into the vector unit, ii) calculate the current contribution for the $n_V$ virtual particles and iii) accumulate this current in the global electric current grid. This last step needs to be handled differently as more than one of the virtual particles may be in the same simulation cell, and there would be a memory conflict when accumulating the results on the global grid. This can just be done serially by looping through each of the $n_V$ virtual trajectories and depositing the previously calculated current for each one. To improve performance and benefit from SIMD calculations in this step, we accumulate all current components in a single vector step. We convert the 3 vectors holding the $x$, $y$ and $z$ electric current components of all $n_V$ virtual particles, into $n_V$ vectors holding all current components (and a dummy value) for each individual virtual particle. We then loop over these vectors and accumulate them in global memory using a vector operation, provided the electric current grid is organized as described in the previous section.

\section{OASCR Joule Metric}
\label{sec_oascr}

\Table{\label{tab_oascr}OASCR Joule Metric test problems. All simulations ran on $\sim 55$ thousand cores. Results are shown for the baseline and improved versions of the code.}
\br
Run&Grid&Particles&\centre{2}{Performance [G part/s]}\\
\ns
& & &\crule{2}\\
       &       &              &Baseline&Improvements\\
\mr
Warm Plasma&$6144 \times 6144 \times 1536$& $4.46 \times 10^{11}$&76.2 & 179.9\\
LWFA 1        &$8064 \times 480 \times 480$    & $3.72 \times 10^{9}$&4.22 & 29.6\\
LWFA 2        &$8832 \times 432 \times 432$    & $6.59 \times 10^{9}$&3.72 & 27.43\\
LWFA 3        &$4032 \times 312 \times 312$    & $1.26 \times 10^{10}$&8.85 & 61.2\\
\br
\endTable

The performance of the new features implemented for real laser wakefield problems were benchmarked within the framework of the 2011 OASCR Joule Software Effectiveness Metric project, that aims to analyze/improve applications requiring high capability/capacity HPC systems. These tests focused on the Jaguar supercomputer at ORNL, a Cray XT5-HE system with a measured performance of 1.76 Pflop/s, that ranked number 3 in the Top 500 list \cite{top500} at the time of the tests. Jaguar had 18680 computing nodes with 2 AMD 6276 cpus with 6 cores each for a total of 224160 computing cores. Each core had an SSE SIMD unit capable of performing 4-wide vector operations in single precision. The project started with a set of initial tests that used $\sim 1/4$ of the full systems. These tests aimed to establish a baseline performance for the code for normal simulation parameters relevant for current research, and allowed us to find the relevant bottlenecks and performance hotspots. The developments described above were then implemented in the code, and the same tests were repeated to measure the performance improvements. At the end of the project  we also analyzed the strong scaling of the test problems using the full system.

The simulation parameters for the tests performed can be found in table \ref{tab_oascr}. To test the baseline performance of the code on modern hardware a uniform warm plasma test, using a temperature distribution with a generalized thermal velocity of $u_{th}Ê= \gamma \beta_{th} =Ê0.1$. The laser wakefield scenarios modeled focused on propagating a 200 TW (6 Joule) laser in a $1.5Ê\timesÊ10^{18}Ê\rm{cm}^{-3}$ plasma with and without moving ions (LWFA-1 and LWFA-3 respectively), and a 1 PW (30 Joule) laser in a $0.5Ê\timesÊ10^{18}Ê\rm{cm}^{-3}$ (LWFA-2).  All simulations were done with quadratic (2nd order) interpolation on a parallel partition of $\sim 55$ thousand cores using the standard fortran version of the code. The measured baseline performance of the code can be found in table \ref{tab_oascr}. The code had a peak performance of 76.2 G particles/s for the warm plasma. However, due to the load imbalance problems mentioned above the performance of the LWFA runs was 9 - 20 times lower.

The same tests were then repeated using a SIMD (SSE) accelerated version of the code on the same parallel partition. The warm plasma test used the pure MPI distributed memory parallel algorithm, and the LWFA runs were done using the hybrid MPI/OpenMP shared memory algorithm, using 6 threads per MPI process (1 MPI process per CPU). The results are summarized in table \ref{tab_oascr}. With the new features the peak performance of the code improved to 179.9 G particles/s, and all LWFA simulations witnessed speedups of $\sim 7$, without any change in the hardware. The gap between peak performance and LWFA performance went down by a factor of $\sim 3$. This speedup comes from a combination a factors. For the LWFA-1 test the breakdown of this speedup was as follows: i) the change in parallel partition, using more nodes longitudinally lowered the imbalance by a factor of 1.91, ii) the shared memory algorithm further lowered the imbalance by a similar factor, improving performance by 1.82 and iii) the SSE code and minor memory optimizations provided an additional 2.02 fold speedup. This behavior was similar for the other LWFA simulations.

\Table{\label{tab_oascr_final}Code performance on the full Jaguar system ($\sim 220$ thousand cores), showing the performance of the PIC algorithm, the speedup from the baseline tests, and the floating point (FP) performance. The HPL benchmark FP performance is also included.}
\br
Run& PIC& Speedup & FP\\
\ns
       & [ G part / s ]    &  & [T Flops]\\
\mr
Frozen (s1)& 1463.5 & - & 516.9 \\
Frozen (s2)& 784.0 & - & 736.1 \\
Warm plasma (strong)& 719.8 & $9.45 \times$ & 679.7\\
Warm plasma (weak)& 741.2 & $9.73 \times$ & 700.1 \\
LWFA 1 (strong) & 70.9 & $16.80  \times$ &76.55\\
HPL & - & - &1759 \\
\br
\endTable

The final tests aimed to analyze code performance in the full system, and to establish our capability in scaling the test problems in this hardware partition. A series of tests were chosen to establish theoretical peak code performance, strong (fixed problem size) and weak (increasing problem size with number of cores) scaling of well balanced problems, and strong scaling of one of the laser wakefield problems. The hardware partition used was exactly 4 times the size of the partition used previously, and corresponds to 98.7\% of the full system. Table \ref{tab_oascr_final} summarizes the performance results for these runs. The frozen plasma tests measure the asymptotic limit for code performance by modeling a zero temperature plasma with $1.86 \times 10^{12}$ particles, using linear (s1) and quadratic (s2) interpolation. In this scenario there is no particle motion, which leads to optimal performance in current deposition and minimizes the communications for particle boundaries. For linear interpolation the code performance was $1.46 \times 10^{12}$ particles / s and for quadratic interpolation $0.78 \times 10^{12}$ particles / s. The warm plasma tests focus on a more realistic scenario that uses a warm plasma distribution that more accurately mimics typical simulation parameters, while maintaining a balanced load. Two scaling tests were made, for strong and weak scaling of this simulation. The total number of particles in these simulations were $4.64 \times 10^{11}$ and $1.86 \times 10^{12}$ particles respectively. The code performance was very similar in both cases, on the order of $0.73 \times 10^{12}$ particles / s, slightly below the frozen plasma limit. In both scenarios the code was found to hyper-scale from the previous tests, with speedups of 9.48 to 16.8, with only 4 times the number of cores. This is as result of both the shared memory algorithm, that improved load balance issues, and the vector optimizations, that improved floating point efficiency. Table \ref{tab_oascr_final} also shows the floating point (FP) performance (floating point operations per second) of the code for the large scale tests. For comparison the High Performance Linpack (HPL) benchmark, which is the standard HPC system performance benchmark \cite{top500}, was included for this system. The code reaches very high floating point / parallel efficiency for the full system, reaching 0.74 PFlops peak performance or 42\% of the HPL benchmark. Quadratic Interpolation was found to improve operation count / memory ratio which offsets the overhead of memory access, achieving higher floating point efficiency. As a result, although the number of operations involved in the quadratic interpolation algorithm is $\sim 2.7 \times$ larger than for linear interpolation, the code is only $\sim 1.8 \times$ slower.

\begin{figure}
\begin{center}
\includegraphics[width=10cm]{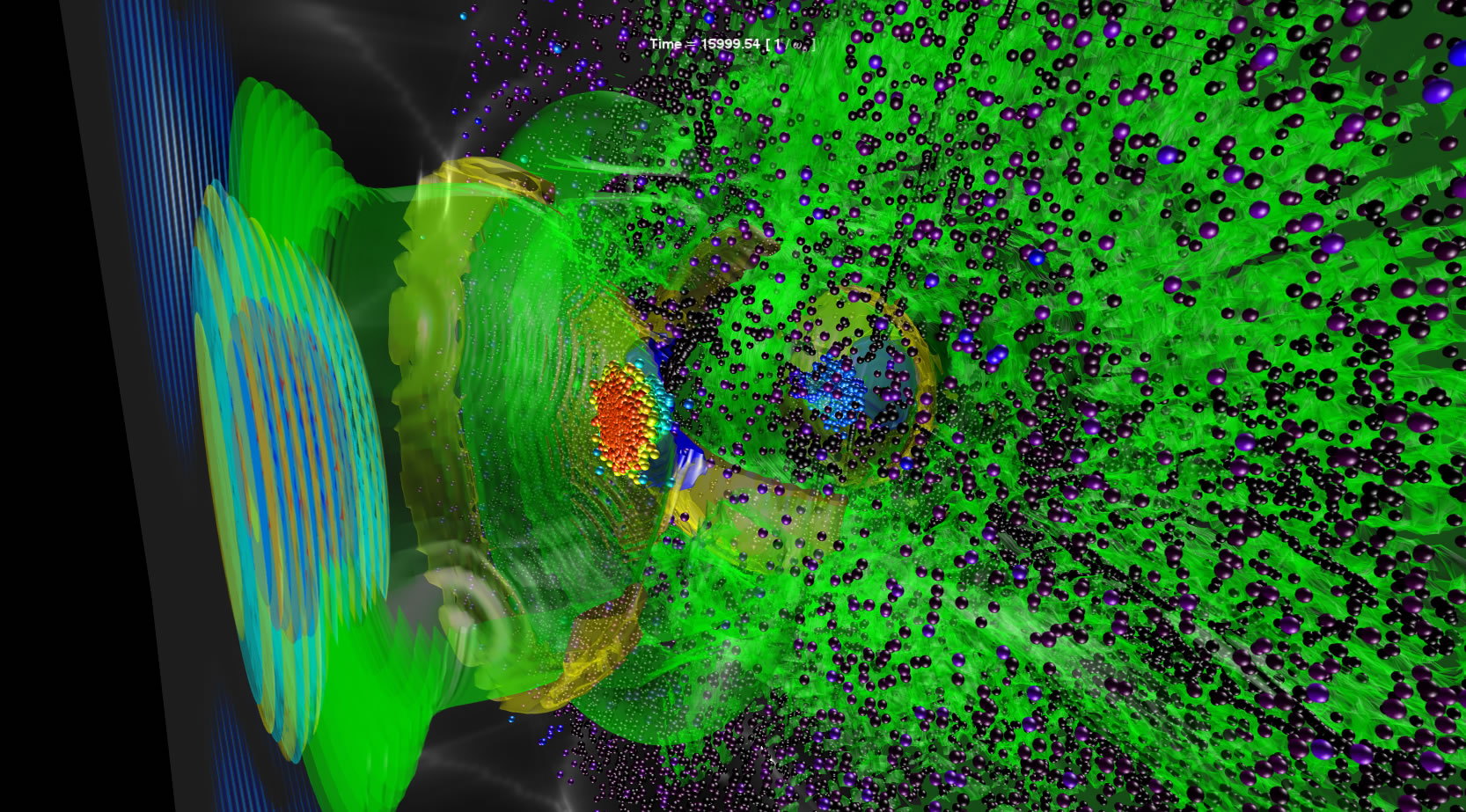}
\caption{Large scale modeling of LWFA showing the laser driver, plasma wave, and accelerated particles (spheres). The particle colors are scaled according to energy, from lowest (black) to highest (red).}
\label{fig_lwfa_strong}
\end{center}
\end{figure}

Finally, figure \ref{fig_lwfa_strong} shows a visualization of the results from the strong scaling of the LWFA  run 1. The global performance of this simulation was $7.09 \times 10^{10}$ particles / s, or $3.12 \mu s$ core push time, which corresponds to a speedup of 16.8 from the previous tests, and a 2.4 speedup from the improved results at 55 thousand cores. The load imbalance was 4.66, larger than for the 55 thousand core partition tests, which is expected since the node volume decreased by a factor of 4. The floating point performance of the code was 77 TFlops (4.4\% of the HPL benchmark). When compared with the warm plasma tests we see that the code ran 9.8 times slower. This is mainly due to the imbalance shown above and the fact that we have a very small final problem size, on average ~ 15 thousand particles per node. In this situation the overhead of the remaining code structure and the parallel communication overhead become important, and lead to a slowdown of about a factor of two, even for a perfectly load balanced situation. These numbers indicate that the LWFA run 1 simulation described above, that would take 2 days and 18 hours using the performance established by the initial tests, can now be performed in under 4 hours using the full Jaguar system.

\section{Performance on Tier-0 Systems}
\label{sec_performance}

To test the baseline performance of the code on modern hardware we chose the same uniform warm plasma test described in the previous sections. We chose a warm plasma instead of a frozen plasma for these tests to benchmark a more realistic scenario where particles move between nodes and stress memory access issues. Table \ref{tab_pic_perf} shows the performance measured on a single workstation with a single 8 core Intel E5-2680 cpu running at 2.7 GHz for a $128^3$ grid with 8 particles per cell (16 million particles total), using all cores. The test used the AVX vector unit of the cpus operating in single precision. Results are shown as a function of different interpolation level orders, from linear (1st order) to quartic (4th order), per core and per cpu. For linear interpolation the code achieved a performance of 134.4 million particle pushes per second per cpu, completing one iteration in the test node in $\sim 60$ ms. The floating point efficiency was about $\sim 50\%$ of peak. 

\Table{\label{tab_pic_perf}OSIRIS performance on an Intel E5-2680 cpu running at 2.7 GHz as a function of interpolation level per core and per cpu.}
\br
Interpolation&\centre{2}{Push time}&\centre{2}{Performance}\\
Level&\centre{2}{[ns/part]}&\centre{2}{[M part/s]}\\
\ns
&\crule{2}&\crule{2}\\
&core&cpu&core&cpu\\
\mr
1& 59.5& 7.4&16.79&134.34\\
2&108.7&13.6&9.20&73.62\\
3&198.3& 24.8&5.04&40.34\\
4&492.2&61.5&14.4&16.25\\
\br
\endTable

\begin{figure}
\begin{center}
\includegraphics[width=8cm]{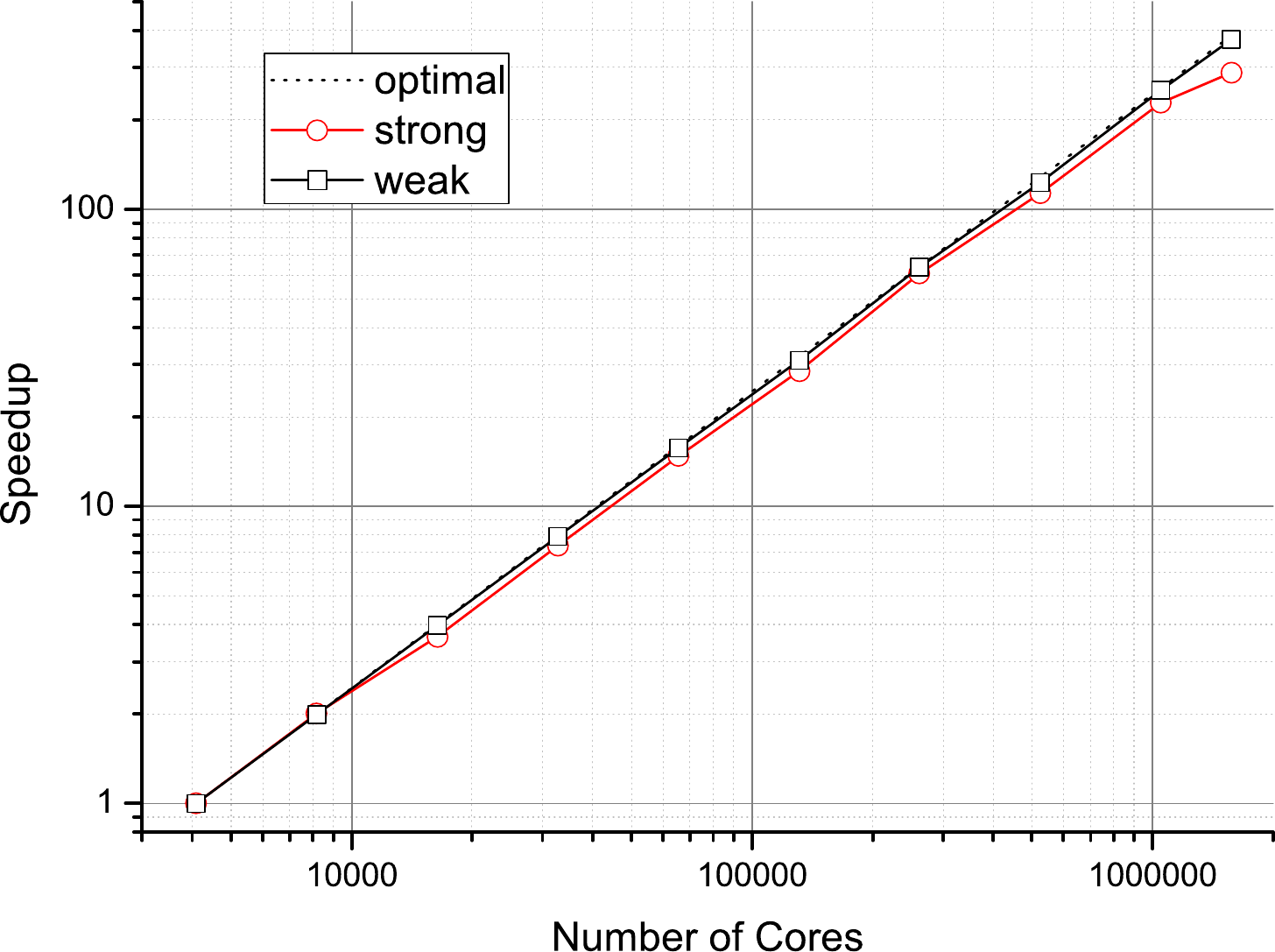}
\caption{Parallel scalability on the Sequoia system at Lawrence Livermore from 4096 to 1572864 cores (full system).}
\label{fig_sequoia}
\end{center}
\end{figure}

We also performed parallel scalability tests at the Sequoia system (IBM BlueGene/Q) at the Lawrence Livermore National Laboratory in the US, which is currently the largest CPU based system in the world, with a total of 1572864 of computing cores. In these tests we analyzed the code scalability starting from a 4096 core partition all the way to the full system partition. We used the same warm plasma test described above with quadratic (2nd order) particle interpolation, and tested both weak parallel scaling and strong parallel scaling. For weak scaling, we used a grid size of  $256^3Ê\timesÊ( N_{\rm{cores}}Ê/Ê4096 )$ and 8 particles per cell, with a final problem size of $\sim 6.4$ Gcells. For strong scaling we chose a problem size of $2048^3$ grid cells with 16 particles per cell. The results are summarized in figure \ref{fig_sequoia}. The weak scaling results show near perfect scaling up to the full system partition, with scaling efficiencies over 96\% for all partitions tested. The choice of a finite difference field solver for the algorithm means that all calculations are local, and each partition elements needs only to communicate with its nearest neighbors, allowing for very high weak scalability, that should also hold for larger systems. The strong scaling tests show a final scaling efficiency of 75\%. This small drop from optimal efficiency comes mainly from 2 factors, namely i) the problem size could not be divided evenly across all cores at the largest partition sizes, because of the high number of cores involved, which meant that some cores had to deal with a larger computational domain resulting in load imbalance and ii) the final problem size per core was only $\sim 5000$ cells, meaning that the ratio between computation and communication is quite small. The final simulation ran in 44.2 s, compared to the initial $\sim 3.5$ hours at 4096 cores.

We also had access to the BlueWaters system at NCSA Illinois with the goal of demonstrating sustained Petascale performance in PIC plasma simulations. The BlueWaters system is a hybrid cpu / gpu system, with an aggregate peak (theoretical) performance of 11.61 PFlops. The tests were run on the cpu partition only that uses a total of 772 480 cores of AMD 6276 cpus running at 2.3 GHz, again using the AVX optimized version of the code in single precision. We used the same warm plasma test, with a problem size of $38624 \times 1024 \times 640$ ($\sim 2.5 \times 10^{10}$) grid cells, with 400 particles/cell, or $\sim 10^{13}$ particles total, which were, to the best of our knowledge, the largest PIC simulations ever performed. The measured average floating point performance of the code was 2.2 PFlop/s, corresponding to 31\% of the peak theoretical performance of the cpu partitions, clearly demonstrating sustained petascale performance.

\section{Overview}

The present problems in laser wakefield acceleration present a formidable challenge to computational physicists. Present day state of the art machines already provide the necessary computing power for detailed large scale simulations on this field, but efficient use of these systems is required. The developments presented in this paper exploit the multi-scale parallelism available in current HPC systems, allowing us to perform large scale numerical modeling of laser wakefield accelerators. 

The simulation tools presented here can push particles in under 8 ns / particle / cpu using modern hardware. This performance was reached in single precision that allows for twice the number of operations to be performed simultaneously in vector operations. Provided the PIC algorithm is carefully implemented, in particular defining particle positions relative to local mesh vertices, this poses no limitation for LWFA modeling, allowing for stable and accurate large scale simulations. The simulation code can scale efficiently to over $\sim 10^6$ cores, and demonstrates an aggregated performance of over $\sim 2 \times 10^{12}$ particle pushes per second using quadratic interpolation, modeling over $\sim 10^{10}$ cells and $\sim 10^{13}$ particles, operating at a sustained floating point performance in excess of 2 PFlop/s. The excellent parallel scalability attained is a result of the algorithm and communication pattern choices: all calculations are local, and communication only occurs with nearest neighbors. The communication overhead will be independent of the global core count and, provided the problem size per core is above a certain threshold, the ratio between computation and communication will render this overhead negligible, allowing for near perfect weak scaling of well balanced problems on this number of cores. 

For LWFA problems however, load imbalance is a serious issue at high core counts. Dynamically load balancing the simulation by adjusting the position of node boundaries, even considering multi-dimensional scenarios, was not found to improve simulation performance. This was mainly due to the very small simulation volume per core that limits the range of motion for the boundaries, and therefore the possibility for improvement, and the overhead involved in redistributing the computational load. Alternatively, exploiting the shared memory parallelism available in current HPC systems, allows for perfect load balance inside each shared memory node and provide a significant speedup, effectively mitigating this issue and allowing for efficient modeling of these scenarios. The tools presented here open new avenues of research between theoretical/ massive computational studies and laboratory experiments in plasma based particle acceleration, with full scale models, accurate physics, and quantitative high fidelity simulations.

\ack

This work was partially funded by the European Research Council (EU) through the Advanced Grant "Accelerates" (ERC-AdG2010 no. 267841), DOE (US) under grants DE-SC0008491, DE-SC0008316 and DE-SC0007970, and NSF (US) under grants OCI-1036224, PHY-0936266, PHY-0960344 and PHY-0934856. This work was also performed under the auspices of the U.S. Department of Energy by Lawrence Livermore National Laboratory under Contract DE-AC52-07NA27344. Simulation and scaling studies were was done as part of the DOE INCITE and OASCR Joule Metric programs at ORNL and on Blue Waters, on the Sequoia supercomputer at LLNL, and at the Jugene/Juqueen supercomputers at JFZ (Germany) through PRACE (EU). F Fiuza would like to acknowledge the Livermore Computing Center for the access to the Sequoia supercomputer and the LLNL Lawrence Fellowship for financial support. RA Fonseca would like to acknowledge the ISCTE - Instituto Universit\'{a}rio de Lisboa Scientific Awards for financial support. The authors would also like to thank Dr. K. Roche at PNNL and Dr. P. Spentzouris at Fermilab for their support and helpful discussions.

\section*{References}

\bibliographystyle{iopart-num}
\bibliography{fonseca_eps_2013}

\end{document}